\UseRawInputEncoding
\documentclass[
reprint,
superscriptaddress,
amsmath,amssymb,amsart,
aps,
prl,
floatfix,
]{revtex4-2}

\usepackage{amsmath}
\usepackage{amsfonts}
\usepackage{amssymb}
\usepackage{graphicx}% Include figure files
\usepackage{dcolumn}% Align table columns on decimal point
\usepackage{bm}% bold math
\usepackage{braket}% bras and kets
\usepackage{amsthm}% proofs
\usepackage{bookmark} % Add bookmarks to pdf
\usepackage[autostyle]{csquotes}
\usepackage{hyperref}
\usepackage{changes}
\usepackage{color,soul}
\usepackage{lineno}
\graphicspath{{./figs/short}}
\usepackage{orcidlink}

\def\u0{\ensuremath { \hat{\bm{u}}_0}}

\begin{document}

\title{The Orbital Angular Momentum of Azimuthal Spin-Waves}

\author{T. Valet \orcidlink{0000-0002-7775-9395}}
\email[Corresponding author :]{ tvalet@mphysx.com}
\affiliation{Université Grenoble Alpes, CEA, CNRS, Grenoble INP, Spintec, 38054 Grenoble, France}

\author{K. Yamamoto \orcidlink{0000-0001-9888-4796}}
\affiliation{Advanced Science Research Center, Japan Atomic Energy Agency, Tokai, Ibaraki 319-1195, Japan}

\author{B. Pigeau \orcidlink{0000-0003-2331-8436}}
\affiliation{Université Grenoble Alpes, CNRS, Grenoble INP, Institut Néel, Grenoble, France}

\author{G. de Loubens \orcidlink{0000-0001-8096-3058}}
\affiliation{SPEC, CEA, CNRS, Université Paris-Saclay, 91191 Gif-sur-Yvette, France}

\author{O. Klein \orcidlink{0000-0001-9429-5150}}
\email[Corresponding author :]{ oklein@cea.fr}
\affiliation{Université Grenoble Alpes, CEA, CNRS, Grenoble INP, Spintec, 38054 Grenoble, France}

\date{\today}

\begin{abstract}
In the context of a growing interdisciplinary interest in the angular momentum of wave fields, the spin-wave case has yet to be fully explored, with the extensively studied notion of spin transport being only part of the broader picture. Here we report experimental evidence for non-zero magnon orbital angular momentum inside magnetic disk, by resolving the frequency splitting between magnon modes with counter-rotating wavefronts and thereby avoiding formation of azimuthal standing waves. This requires an unambiguous formulation of spin and orbital angular momenta for spin waves, which we provide in full generality taking advantage of a systematic application of quantum field theory techniques. % as detailed in an associated article \cite{Valet2025a}. 
The results unequivocally establish magnetic dipole-dipole interactions as a magnetic-field controllable spin-orbit interaction for magnons. Our findings open a new research direction, leveraging the spectroscopic readability of angular momentum for azimuthal spin waves and other related systems.
\end{abstract}

\hfill \textbf{LC20026}

\maketitle

%\linenumbers

There has been an increasing realization in recent decades of the fundamental importance of the angular momentum (AM) carried by wave fields~\cite{chen:2020,franke:2022}, which can be separated into spin (SAM) and orbital (OAM) components in certain cases. The latter is a universal feature of waves in uniform continuum media represented by helical or rotational wavefronts, and can potentially encode a large amount of information for mode multiplexed communication channels or multi-level registers of quantum states. Theoretical and experimental investigations of wave AM are already well developed for electromagnetic waves~\cite{coullet:1989,allen:1992,hancock:2021}, plasma waves~\cite{bliokh:2022d}, fluid waves~\cite{jones:1973,thomas:2003} or elastic waves~\cite{garanin:2015}. Acknowledging wave AM in solid state media should be termed pseudo AM~\cite{mcintyre:1981,streib:2021}, we omit the prefix following the common practice of the magnetism community. In recent years, experiments have found AM transfer between spin-waves (SWs) and optical vortices~\cite{osada:2016,haigh:2016,zhang:2016,Osada2018PRL,Gloppe2019}, or elastic waves~\cite{an:2020,xu:2020,sasaki:2021,schlitz2022,an:2023,Liao2023}. Most of these works, however, focused on SAM of SWs, leaving their OAM experimentally unresolved. Although Kerr and Brillouin light scattering (BLS) microscopy successfully visualized the azimuthal wavefronts in magnetic vortex states~\cite{Buess2004,Park2005,Vogt2011,Schultheiss2019}, they did not discuss any associated OAM. The situation may also be placed in the context of the general difficulty in directly observing rotating wavefronts~\cite{Leach:2002}. 

In this Letter, we report on a spectroscopic measurement of SW eigenstates with nonvanishing OAM in a normally magnetized disk, paving the way for powerful detection schemes of general wave OAM through the aforementioned transduction of SWs to electromagnetic and elastic waves. This leverages on the lifted degeneracy between the SW modes with counter-rotating wavefronts, which originates from a type of spin-orbit interaction (SOI) that communicates the broken time-reversal symmetry associated with the equilibrium magnetization to SW textures. While a handful of theoretical works discuss AM of magnetization dynamics, none of them addresses the crucial role of SOI. They also have limited scopes, either specifying certain textures~\cite{Ivanov1998,Buess2005,Guslienko2008,Verba2021} and geometries~\cite{sharma:2017,osada:2018,rychly2018,jia:2019,jiang:2020,lee:2022} or not discussing azimuthal SW eigenstates~\cite{tsukernik:1966,goldstein:1984,yan:2013,tchernyshyov:2015,streib:2021}. We shall also mention recent theories of OAM~\cite{neumann:2020,fishman:2022} and SOI~\cite{liu:2020} across the magnon Brillouin zone, which are distinct topics from long wavelength SWs in mesoscopic ferromagnets. As a prerequisite to the spectroscopic detection of AM eigenstates, therefore, we developed a general formulation of magnon AM that clarifies its relation to azimuthal SW eigenstates in any axisymmetric geometry, whose details are exposed in a separate article~\cite{Valet2025a}. We identify the dynamical dipole-dipole interaction (DDI) to be the generic SOI for SWs, and assign correct AM indices to various spectral lines, which was not done properly in the previous studies~\cite{Guslienko2008,osada:2018,Osada2018PRL}. Agreement with the experiment in a fully saturated sate is quantitative, and establishes a spectroscopic measurement of the SW SOI and assignment of OAM quantum number. Our findings lay a foundation for reading OAM states not only for SWs but also for phonons or photons that can hybridize with SWs, and thereby open a new research direction in the study of general wave AM. 

\begin{figure}[t!]
\includegraphics[width=0.45\textwidth]{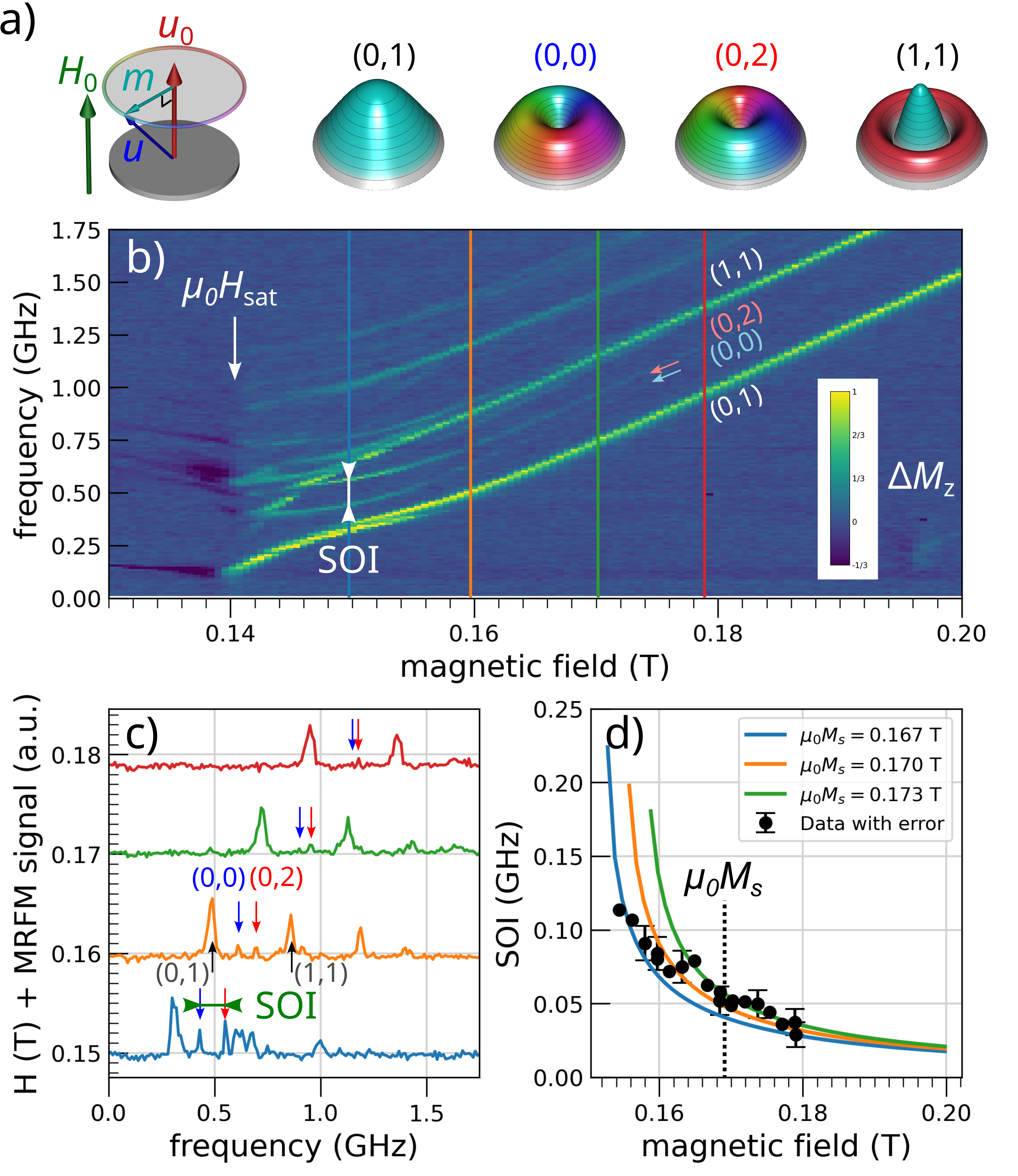}
\caption{\label{fig:1} a) Graphical representation of the precession pattern of ${\bm m} ({\bm x})$ for SW modes labeled $(n_R,n_J)$. b) MRFM (Magnetic Resonance Force Microscopy) spectroscopy as a function of normal magnetic field and frequency on a YIG disk. c) Line cuts at field values indicated by vertical lines of respective colors in b). The split between the $(0,2)$ and $(0,0)$ peaks defines the SOI. d) Magnetic field dependence of the SOI. The dots are the experimental points, while the solid lines are theoretical predictions calculated by Eq.~(\ref{eq:SOI}) for 3 different values of $\mu_0 M_s=$0.167, 0.170, and 0.173~T corresponding to its uncertainty range.}
% \vspace{-10 pt}
\end{figure}

Figure~\ref{fig:1} shows the measured SW spectra near the saturation field, $H_\text{sat}$~\cite{supp}.  The spectroscopy is conducted by a magnetic resonance force microscope (MRFM)~\cite{naletov:2011}, an instrument that employs a soft cantilever with a magnetic nanosphere at its apex to mechanically detect the magnetization dynamics in the sample placed underneath. It probes the variation of the magnetization along the local equilibrium axis, enabling the detection of modes with nonuniform spatial profiles, which are difficult to characterize with conventional probes. The sample is a Yttrium Iron Garnet (YIG) microdisk, which has been patterned by Ar etching from a thin film grown by liquid phase epitaxy (LPE), with a diameter of 1~$\mu$m and a thickness of 55~nm determined from the splitting between the Kittel mode and the first higher order thickness standing spin-wave mode (exchange dominated) observed by BLS, assuming that the exchange length in YIG $\lambda_\text{ex} = 15$~nm~\cite{Beaulieu2018}. Subsequently, a radio-frequency antenna is deposited on its surface. The apparatus is situated between the poles of an electromagnet that produces a large magnetic field, $H_0$. In our setup, a slight misalignment exists in the tilt angle ($<1^\circ$) between the applied magnetic field and the normal of the disk~\cite{supp}. As will be shown below, this tilt seems sufficient to excite nonvanishing OAM SW modes. Deviation of the field from the normal introduces a rigid shift of the whole spectrum for $H_0 \gtrsim M_s$~\cite{supp} and hence should have no impact on analyzing the frequency separation between modes. More details about the experimental setup and the YIG disk can be found in the Supplemental Material ~\cite{supp} (see also references \cite{landolt70,Klingler2014,Makino1981,sangiao17,klein08,Charbois2002,Taurel2016,sangiao17} therein).

\begin{figure}[t!]
\includegraphics[width=0.49\textwidth]{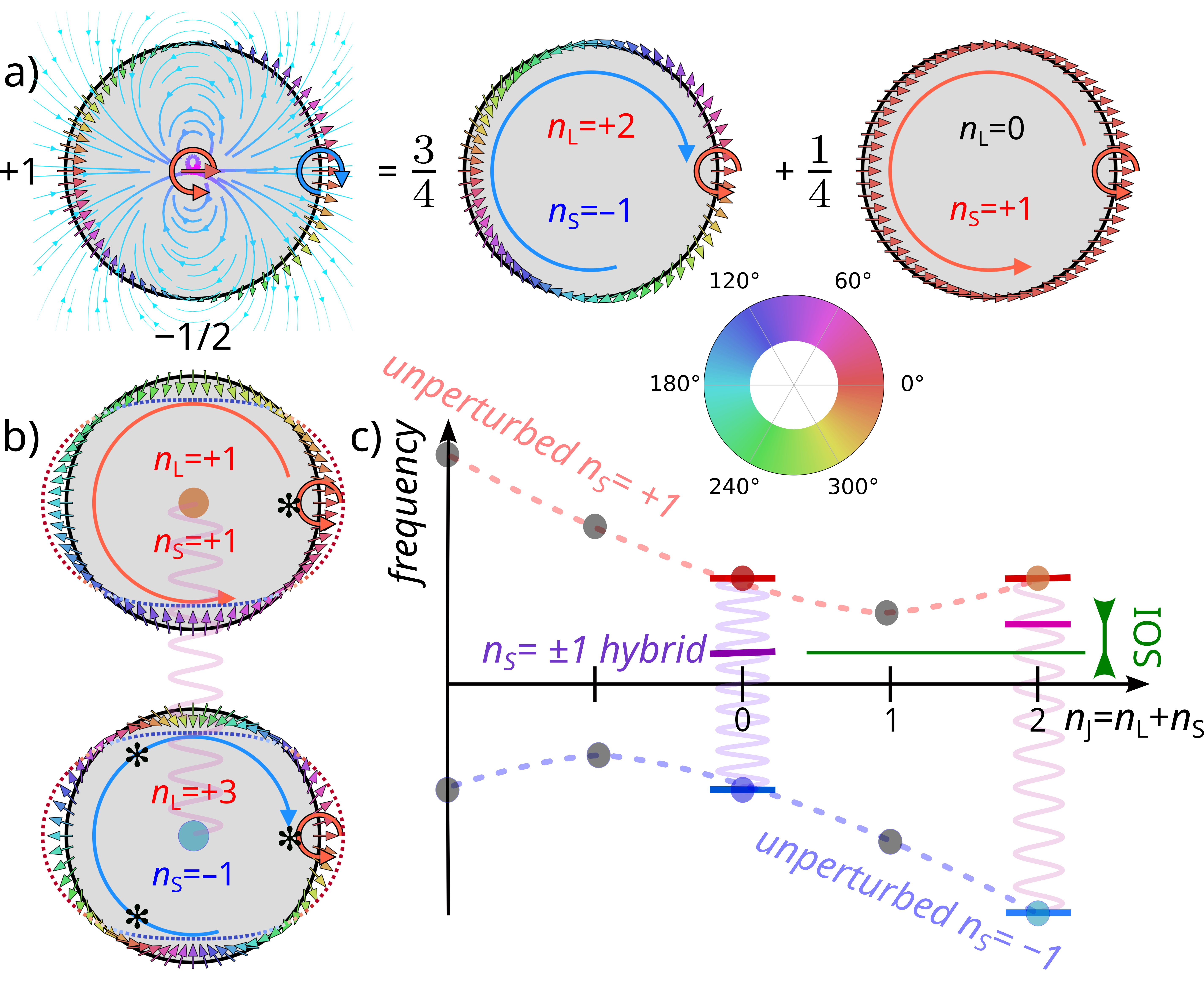}
\caption{\label{fig:2} (a) Illustration of the azimuthal pattern of the dynamical DDI. The stray magnetic field profile at a fixed radius from the counterclockwise rotating magnetic dipole that generates it is a $3:1$ superposition of the $(n_L = +2,\, n_S = -1)$ and $(n_L = 0,\, n_S = +1)$ in the OAM eigenmode basis. (b) Coupling (wavy line) of the two OAM eigenmodes with $n_J =2$ \textit{via} dynamical DDI. The asterisk helps count $n_L$: the repetition of a given orientation (here at $0^\circ$, cf. color wheel). (c) Comparison of the dynamical DDI coupling for two states of opposite OAM $n_L=\pm 1$. The red(blue) dotted line shows the dispersion of the $n_R=0$ eigenstates of $\hat{\mathcal{O}}_0$ for $n_S=+1$($n_S=-1$). The unperturbed $(n_L=\pm 1,n_S=+1)$ eigenstates undergo energy lowering through hybridization mediated by $\hat{\mathcal{O}}_d^{(-)}$ with $n_S=-1$ states of different $|n_L|$. This gives rise to two distinct elliptically precessing states (magenta levels) depending on the polarity of $n_L$. The frequency shifts are inversely proportional to the gap between the $n_S =\pm 1$ branches, resulting in a SOI splitting that increases when the applied magnetic field approaches saturation.
}
\vspace{-10 pt}
% \end{figure*}
\end{figure}

Figure~\ref{fig:1}b) is a density plot of the variation of the MRFM signal as a function of $H_0$ and the excitation frequency, $\omega/(2 \pi)$. The spectral peaks are labeled by radial index $n_R \in \mathbb{Z}^+ $ and azimuthal index $n_J \in \mathbb{Z}$, to be discussed in detail shortly. In this plot, the most intense signals arise from the Kittel mode $(n_R ,n_J)=(0,1)$, and its first radial harmonic $(1,1)$. Expected precession patterns of their dynamical magnetization $\bm{m}({\bm x})$ are sketched in Fig.\ref{fig:1}a) where the height indicates the amplitude and the color the azimuthal orientation (cf. colorwheel in Fig.~\ref{fig:2}). Because of their spatially uniform precession along the azimuthal direction, these are naturally the modes coupled strongest to the antenna. By linearly fitting the variation of the $(0,1)$ mode frequency with the external magnetic field between 0.3 and 1~T, we determine the values of the gyromagnetic ratio $|\gamma | = (1.77 \pm 0.01)  \times 10^{11}$~$\text{rad} \cdot \text{sec}^{-1} \cdot \text{T}^{-1} $. The saturation magnetization $ \mu _0 M_s = 0.170 \pm 0.003$~T ($\mu _0$ is the vacuum permeability) is extracted from the ferromagnetic resonance of the extended thin film with the magnetic field applied in-plane~\cite{supp}.  The obtained value is in agreement with previous studies made on LPE grown YIG thin films~\cite{Beaulieu2018}. Between these landmark eigenstates, we also observe two additional small peaks which can be uniquely assigned to the modes $(0,0)$ and $(0,2)$. No other modes are expected in this frequency range given our disk dimensions. These two eigenstates correspond to SW modes with counter-rotating wavefronts, as illustrated in Fig.~\ref{fig:1}a). Their split in frequency increases as $H_0$ approaches $H_\text{sat}$ from above. We will establish below that this is due to dynamical DDI, whose influence increases with decreasing $H_0$. It introduces an ellipticity in the spin precession of the homogeneously magnetized disk and acts as a SOI for SW. The field-controllable SOI can be made much larger than the linewidth by tuning $H_0$ closer to $H_{\rm sat}$, making the split readily observable.

To substantiate the claims made in the previous paragraph, we set up a theoretical model, relate SW eigenstates to OAM, and compute their spectrum. Consider a ferromagnet occupying a region $\Omega$, whose magnetization is given by \smash{${\bm M}({\bm x},t) = M_s {\bm u}({\bm x}, t)$}, with \smash{${\bm u}({\bm x}, t)$} a time dependent unit vector field (Fig.~\ref{fig:1}a)). We assume the existence of a textured equilibrium, \smash{${\bm u}_0({\bm x})$}, and define the dimensionless SW field \smash{${\bm m} = {\bm u} - {\bm u}_0$}. The linear dynamics of $\bm{m}$ is determined by minimizing the action $\mathcal{S}=\mu _0 M_s^2 \int dt \int _{\Omega }d^3 x~\mathcal{L}\equiv \mu _0 M_s^2 V\int dt \langle \mathcal{L}\rangle _{\Omega }$, where $V$ the finite volume of $\Omega $, $\langle \cdot \rangle _{\Omega }$ denoting spatial averaging, and the normalized Lagrangian density $\mathcal{L}$ is given by~\cite{tsukernik:1966,Valet2025a} 
\begin{equation} \label{eq:lagrangian_tot}
	\mathcal{L} = \frac{1}{2} \frac{\left( {\bm u}_0 \times {\bm m} \right)}{\omega_M} \cdot \partial_t {\bm m} - {\mathcal U} - \kappa \ {\bm u}_0 \cdot {\bm m} . \\
\end{equation}
The first term plays the role of a kinetic energy density, with $\omega_M = |\gamma| \mu_0 M_s$ a natural frequency scale. As for the "potential" energy density~\cite{brown:1963,miltat:2002}, we choose for concreteness ${\mathcal{U}=\frac{1}{2}h_0 \bm{m}\cdot \bm{m} + \frac{1}{2}\lambda _{\rm exc}^2 \nabla\bm{m}:\nabla\bm{m} -\frac{1}{2}h_a (\bm{e}_a \cdot \bm{m}) ^2  +\mathcal{U}_d} $, where $h_0 \bm{u}_0$ is the equilibrium effective field, $\lambda _{\rm exc}$ the exchange length, $h_a$ the anisotropy field along the unit vector $\bm{e}_a$, and $\mathcal{U}_d =-\frac{1}{2}(\nabla \phi )^2 +\bm{m}\cdot \nabla \phi $ the dynamical DDI contribution with $\phi $ the magnetic scalar potential. The fields $h_0$, $h_a$, and $-\nabla \phi $ shall be understood as having been normalized by $\mu _0 M_s$. Finally, $\kappa({\bm x}, t)$ is a Lagrange multiplier. 
AM is the conserved quantity associated with a rotation symmetry of the Lagrangian. Let $\delta r_J$ be the infinitesimal rotation by an angle $\delta \theta $, around a certain $Oz$ axis. It admits a decomposition \smash{$\delta r_J = \delta r_L \circ \delta r_S$} where the orbital $\delta r_L$ and spin $\delta r_S$ parts act on $\bm{x}$ and $\bm{m}$ respectively. We introduce a global cylindrical coordinate system $(O;r,\theta,z)$ with local unit vector basis $\{ \bm{e}_r , \bm{e}_{\theta }, \bm{e}_z \}$. The action of $\delta r_J$ on $\bm{m}(\bm{x}) $ is then given by $\delta r_J : \bm{m}\rightarrow \bm{m} + \delta \bm{m}_J $ and $\frac{\delta \bm{m}_J}{\delta \theta } \equiv \frac{\delta \bm{m}_L}{\delta \theta }+\frac{\delta \bm{m}_S}{\delta \theta } =-\partial _{\theta }\bm{m} + \bm{e}_z \times \bm{m}$. If we assume the Lagrangian to be invariant under \smash{$\delta r_J$}, \smash{$\delta r_L$} or \smash{$\delta r_S$}, an application of the Noether's theorem \cite{weinberg:1995} yields respectively, as globally conserved quantities, the projections along $Oz$ of the volume integrated AM, $J^z$, OAM, $L^z$, and SAM, $S^z$, as \cite{Valet2025a} 
\begin{subequations} \label{eq:momenta}
\begin{eqnarray}
    %J^z & = & \mu_0 M_s^2  \frac{\partial {\mathcal L}}{\partial (\partial_t {\bm m})} \cdot   \frac{\delta {\bm m}_J}{\delta \theta} = L^z + S^z,\label{eq:total_momentum} \\
    J^z & = & L^z + S^z,\label{eq:total_momentum} \\
    % L^z  & = & - \frac{\mu _0 M_s^2}{2 \omega_M} ({\bm u}_0 \times {\bm m}) \cdot \partial_\theta {\bm m } , \label{eq:orbital_momentum} \\
    L^z  & = & - \mathcal{J}_{\scriptscriptstyle M} \int_\Omega d^3 x ({\bm u}_0 \times {\bm m}) \cdot \partial_\theta {\bm m } , \label{eq:orbital_momentum} \\
    % S^z & = & + \frac{\mu _0 M_s^2}{2 \omega_M} \left( {\bm u}_0 \times {\bm m} \right) \cdot  \left( {\bm e}_z \times {\bm m} \right) . \label{eq:spin_momentum}
    S^z & = & + \mathcal{J}_{\scriptscriptstyle M} \int_\Omega d^3 x \left( {\bm u}_0 \times {\bm m} \right) \cdot  \left( {\bm e}_z \times {\bm m} \right) , \label{eq:spin_momentum}
\end{eqnarray}
\end{subequations} 
in which \smash{$\mathcal{J}_{\scriptscriptstyle M} = M_{\scriptscriptstyle S} / (2 |\gamma| )$} emerges as a natural scale of AM density for the linear SWs.

We next clarify the relation between the AM densities and azimuthal SW eigenmodes. The Euler-Lagrange equation for $\bm{m}$ associated with $\mathcal{S}$ reads \smash{$\partial_t {\bm m} = \omega_M \ {\bm u}_0 \times \frac{\delta {\mathcal U}}{\delta {\bm m}}$}, under the constraint \smash{${\bm u}_0 \cdot {\bm m} = 0$,} which we recognize as the linearization of the Landau-Lifshitz equation~\cite{landau:1935}. The Fourier transform in time \smash{${\bm m}({\bm x}, t) = \Re [ \tilde{\bm m}({\bm x}) e^{- {\bf i} \omega t}]$} with $\omega  \in {\mathbb R}$ and \smash{$\tilde{\bm m} \in {\mathbb C}^3  /  \{ {\bm u}_0 \cdot \tilde{\bm m} = 0 \} $} yields an eigenvalue problem $\omega \tilde{\bm m} = \hat{\mathcal{O}} \tilde{\bm m}$, where the concrete expression of the integro-differential operator $\hat{\mathcal{O}}=\bm{u}_0 \times \frac{\delta ^2 \mathcal{U}}{\delta \bm{m}\otimes \delta \bm{m}}$ is given in Eq.~(55) of the accompanying paper~\cite{Valet2025a}. The resultant SW eigenmodes, labelled by a stand-in mode index $\nu $ to be specified by boundary conditions, satisfy two general properties~\cite{mills:2006,daquino:2009,naletov:2011} : (i) if $\lbrace\omega_\nu, \tilde{\bm m}_\nu\rbrace$ is a SW eigenpair, then $\lbrace-\omega_\nu, \tilde{\bm m}_\nu^*\rbrace$ is one as well, representing the same physical state, and (ii) the SW eigenmodes satisfy the orthogonality relation \cite{naletov:2011}
\begin{equation} \label{eq:sw_ortho}
\vspace{- 2 pt}
	 -\frac {\bf i}  2 \frac{\langle \left( {\bm u}_0 \times \tilde{\bm m}_\nu^{*} \right) \cdot \tilde{\bm m}_{\nu'} \rangle_{\Omega}}{{\rm sgn} (\omega_\nu) }  =   A_\nu^2 \ \delta_{\nu, \nu'} ,  
\vspace{- 2 pt}
\end{equation}
with \smash{$ A_\nu  > 0$} being {\em defined} as the natural norm of the mode.
%~\footnote{One could loosely interpret $A_\nu \approx \sqrt{1 -\cos \theta}$, with $\theta \ll 1$ being the spatially averaged cone angle of the magnon.} 
If the action is invariant under $\delta r_J$, the eigenmodes can be chosen as eigenfunctions of the associated infinitesimal generator {\em i.e.}, \smash{${\bm m}_{\nu, n_J} = \Re \left[ \tilde{\bm m}_{\nu, n_J}(r, z) e^{{\bf i} ( n_J \theta - \omega_{\nu, n_J} t)} \right] $} with \smash{$\tilde{\bm m}_{\nu, n_J} \equiv \tilde{m}^r_{\nu ,n_J}(r,z) \bm{e}_r  +\tilde{m}^{\theta }_{\nu ,n_J}(r,z) \bm{e}_{\theta }+\tilde{m}^z_{\nu ,n_J}(r,z)\bm{e}_z $} and $n_J \in \mathbb{Z}$ being an eigenvalue of $\delta r_J$, since \smash{$\left[ \left( {\bm e}_z \! \times \cdot \right) - \partial_\theta \right] \tilde{\bm m}_{\nu, n_J} \! = \! { - {\bf i} n_J \tilde{\bm m}_{\nu, n_J}}$}. Substituting this waveform into Eq.~(\ref{eq:total_momentum}) gives
\begin{equation}\label{eq:total_AM_averaged}
J^z_{\nu, n_J} = {\rm sgn}(\omega_{\nu, n_J})  n_J V \mathcal{J}_M  A_{\nu ,n_J}^2.
\end{equation}
If the action is additionally invariant under $\delta r_L$ and $\delta r_S$, the eigenmodes can be eigenfunctions of both associated infinitesimal generators {\em i.e.}, \smash{${\bm m}_{\nu, n_L, n_S}^{0}= m_{\nu, n_L}^{0}(r, z) \ \Re [ ( {\bm e}_r + {\bf i} n_S {\bm e}_\theta ) e^{{\bf i} ( n_J \theta - n_S \omega_{\nu, n_L}^{0} t)} ]$,} with the corresponding eigenvalues \smash{$n_L = n_J - n_S$}, \smash{$n_L \in {\mathbb Z}$}, \smash{$n_S = \pm 1$} and \smash{$\omega_{\nu, n_L}^0 > 0$ ;} as \smash{${\bm e}_z \times ( {\bm e}_r \pm {\bf i} {\bm e}_\theta ) = \mp {\bf i} ( {\bm e}_r \pm {\bf i} {\bm e}_\theta )$}. Then Eqs.~(\ref{eq:orbital_momentum}-\ref{eq:spin_momentum}) give \smash{$ L_{\nu, n_J}^z = n_S n_L V \mathcal{J}_M A_{\nu, n_J}^2$,} and \smash{$ S_{\nu, n_J}^z = V \mathcal{J}_M A_{\nu, n_J}^2 $}. The canonical quantization for constrained systems~\cite{dirac:1950,dirac:1964,Valet2025a} establishes $V \mathcal{J}_M A_{\nu ,n_J}^2 =\hbar $ for single magnon states, confirming the quantization of conserved angular momenta per magnon.

% \begin{figure*}[ht!]

Let us elaborate on the physical meaning of the angular momentum labels $n_J$ and $n_L$. The Lagrangian is generically invariant under $\delta r_J$ if : (i) the region $\Omega$, (ii) the equilibrium texture ${\bm u}_0$, and \mbox{(iii) all} the magnetic material properties, are invariant under $\delta r_J$. This defines a broad class of axisymmetric ferromagnets, such as YIG spheres~\cite{osada:2016,sharma:2017} or mesoscopic disks in one of their known axisymmetric equilibria~\cite{guslienko:2008b,taurel:2016,vukadinovic:2011,rohart:2013}.
%~\footnote{Even though the interfacial Dzyaloshinskii-Moriya interaction necessary to stabilize a skyrmion state is not included in our present model of micromagnetic energy, it can easily be added and it will not break the cylindrical symmetry.}, 
According to Eq.~(\ref{eq:total_AM_averaged}), $n_J$ is a measure of AM up to the choice ${\rm sgn}(\omega _{\nu ,n_J})=\pm 1$. In all concrete calculations, we use the positive sign convention, by which $n_J >0$ represents positive AM. OAM can be associated to the integer label $n_L$ \emph{only if} the Lagrangian is invariant under $\delta r_L$, which occurs if, in addition to (i-iii), we also \mbox{have :} (iv) ${\bm u}_0 = {\bm e}_a = {\bm e}_z$, and (v) a negligible dynamical DDI~\cite{tsukernik:1966,goldstein:1984},~\footnote{It shall be noted that a negligible dynamical DDI is a {\em sufficient} condition for magnons to carry separately conserved and quantized SAM and OAM, but {\em not} a necessary condition.
%For instance, some families of Walker's modes in saturated spheroids are exact OAM and SAM eigenmodes, while the DDI is often a dominant energy term in that case.
}. Hereafter, we further specialize the problem to a sufficiently thin disk so that $\bm{m}$ can be considered independent of $z$. Then, under (i-v), the equation $\omega \bm{m}=\hat{\mathcal{O}}\bm{m}$ reduces to
\begin{equation}
 \omega^0_{\scriptscriptstyle n_R, n_{L}, n_{S}} m_{\scriptscriptstyle n_R, n_L}^0 \! \! = n_S \left[ h_0  + \lambda_\text{exc}^2 \left( \frac{n_L^2}{r^2} - \nabla^2_\perp \right) \right] m_{\scriptscriptstyle n_R, n_L}^0 , \label{eq:noSOI}
\end{equation}
in which \smash{$\nabla^2_\perp \equiv \partial_r^2 + (1/r) \partial_r$} and $n_R$ is the associated radial index. It immediately follows that \smash{$\omega^0_{\scriptscriptstyle n_R, n_{L}, n_{S}} = n_S \omega^0_{\scriptscriptstyle n_R, |n_{L}|}$} with \smash{$\omega^0_{\scriptscriptstyle n_R, |n_{L}|} > 0$}, while the corresponding eigenvector amplitudes can be expressed as \smash{$\tilde{\bm{m}}^0_{n_R , n_L, n_S} = m_{n_R , |n_L|}^0( r,z) (\bm{e}_r + {\bf i} n_S \bm{e}_{\theta })$}. Similarly to $n_J$, $n_L$ quantifies OAM only up to the sign of eigenfrequency, or equivalently the eigenvalue $n_S$ of $\delta r_S$ that carries no physical significance. In Fig.~\ref{fig:2}b), the AM and OAM of the $n_J=2$ SW eigenstates derived from Eq.~(\ref{eq:noSOI}) are shown schematically. The dotted line highlights in red/blue the number of oscillatory cycles $|n_J |$ of the radial component of $\bm{m}$ around the disk circumference. It differs by one unit from $|n_L |$, which counts the number of revolutions of $\bm{m}$, encoded in the cyclic color wheel. For $n_L \neq 0$ and fixed $n_S$, the sign of $n_L$ indicates whether the direction of rotation of the Cartesian component wavefront (large circular arrow) is identical or reversed compared to the direction of local precession (small circular arrow) \emph{i.e.}, whether the OAM is parallel or anti-parallel to the SAM.

We now introduce the general notion of magnon SOI. 
Consider a system where (i-v) are satisfied and fix $n_L \in \mathbb{Z}$. The magnons come in degenerate pairs $n_J = \pm n_L+n_S$ with identical radial profile and opposite values of the OAM $\propto \pm n_S n_L$. The SOI manifests itself through the lift of this degeneracy, and is quantified by the frequency splittings between magnons with opposite OAM {\em i.e.}, \smash{${\rm SOI}_{n_R, n_L} = (\omega_{n_R, 1 + n_L} - \omega_{n_R, 1 - n_L})$} under the convention $n_S =+1$. In the present case, the splitting is induced by dynamical DDI. When neglecting the dynamical DDI, all the SW eigenmodes of the disk are circularly polarized with right-handed precession, due to the time reversal symmetry breaking associated with the equilibrium magnetization as shown in Fig.~\ref{fig:2}b). However, they induce a dynamic dipolar field with both right- and left-handed circularly polarized components in the disk plane [Fig.~\ref{fig:2}a)]. More quantitatively, the linear operator $\hat{\mathcal{O}}$ can be split as \smash{$\hat{\mathcal{O}} = \hat{\mathcal{O}}_0 + \hat{\mathcal{O}}_d^{(+)} + \hat{\mathcal{O}}_d^{(-)}$} with $\hat{\mathcal{O}}_0$ corresponding to the equilibrium effective field and exchange contributions. $\hat{\mathcal{O}}_d^{(\pm )}$ are defined to be parts of the dynamical DDI that commute and anti-commute with $\delta r_S$ respectively. Therefore, represented in the basis $\{ \tilde{\bm{m}}^0_{n_R ,n_L,n_S} , n_L \in \mathbb{Z}, n_S =\pm 1 \} $, $\hat{\mathcal{O}}_0+\hat{\mathcal{O}}_d^{(+)}$ leaves the $n_S =\pm1$ subspaces invariant, while $\hat{\mathcal{O}}_d^{(-)}$ swaps the two and is the sole source of SOI, with $\hat{\mathcal{O}}_d^{(\pm )}\bm{m}^0_{n_R ,n_L ,n_S}$ proportional to right- and left-handed components of the dynamical stray field. They both conserve $n_J =n_L+n_S$, hence $\hat{\mathcal{O}}^{(-)}_d$ uniquely couples $n_L$ with $n_S=+1$ to $n_L +2$ with $n_S=-1$, breaking the $\pm n_L$ spectral symmetry as illustrated in Fig.~\ref{fig:2}c) where the eigenvalues $\omega ^0_{n_R ,n_L, n_S}$ of $\hat{\mathcal{O}}_0$ \emph{i.e.}, ``unperturbed'' frequency spectrum, are plotted for the $n_R=0$ branch with the dotted lines being a guide for the eyes. In some intermediate field range $h_0 \gtrsim 1$, we can try to obtain a semi-quantitative insight into the emergence of the SOI by considering the dynamical DDI as a perturbation. Focusing on the \smash{$n_R = 0$} branch, and up to second order in the dynamical DDI, we obtain~\cite{Valet2025a} \begin{equation}
{\rm SOI}_{0, n_L}  = \sum_{n_R} \left[ \dfrac{{\Gamma^-_{\langle 0|n_R \rangle,n_L}}^2}{2\omega_K + \Delta^-_{\langle 0|n_R \rangle,n_L}} - \dfrac{{\Gamma^+_{\langle 0|n_R \rangle,n_L}}^2}{2\omega_K + \Delta^+_{\langle 0|n_R \rangle,n_L} } \right], \label{eq:SOI}
\end{equation}
where $\Gamma^\pm_{\langle 0|n_R \rangle,n_L} = \langle m^0_{n_R, |2 \pm n_L|} \hat{\mathcal O}_{d}^{\scriptscriptstyle (-)} m^0_{0, |n_L|} \rangle_{\Omega}$ and $\Delta^\pm_{\langle 0|n_R \rangle,n_L} = \omega^0_{0,|n_L|}+\omega^0_{n_R,|2\pm n_L|}-2 \omega_K$ with $\omega _K$ being the frequency of the quasi-uniform Kittel mode for the disk.  By direct inspection of Eq.~(\ref{eq:SOI}) we obtain the confirmation of a field-controllable SOI, with a leading asymptotic term inversely proportional to $\omega_K$. The splitting between $(n_L=\pm 1, n_S=+1)$ through the hybridization with $(n_L=2\pm1,n_S=-1)$ is illustrated in Fig.~\ref{fig:2}c). The eigenfrequencies and eigenmodes of \smash{${\hat{\mathcal O}}_{0}$} can be obtained semi-analytically from a Galerkin projection of the corresponding spectral boundary value problem over a truncated basis of the analytical exchange SWs, while explicit integral forms of the matrix elements of \smash{${\hat{\mathcal O}}_{d}^{\pm }$} can be obtained on the same basis~\cite{Valet2025a}. This in principle enables a numerical evaluation of the approximate SOI frequency split Eq.~(\ref{eq:SOI}). However, it is found more straightforward, and more accurate over the whole field range, to perform a Galerkin projection of the full spectral boundary value problem associated with \smash{${\hat{\mathcal O}}$}~\cite{Valet2025a}. The SOI frequency split can then be directly obtained from the numerically estimated exchange-dipole SW frequencies. This is the method that has been used to generate the continuous line in Fig.~\ref{fig:1}d).

We now compare the semi-analytical prediction with the experiment. In Fig.~\ref{fig:1}d), we plot the variation of the experimentally determined SOI splitting as a function of the magnetic field strength $H_0$. The superimposed solid lines are the theoretically computed~\cite{Valet2025a} ${\rm SOI}_{0,1}$ obtained for three values of $\mu _0 M_s=$ 0.167, 0.170, and 0.173~T corresponding to the uncertainty in the independent measurement of $M_s$. We find quantitative agreement between the data and the model for field strengths $H_0 \geq M_s$. Moreover, the observed experimental splitting exhibits a clear field dependence, which rules out the possibility of an accidental fixed pinning of comparable strength. The deviations observed for $H_0 < M_s$ can be attributed to the details of how the magnetization projection along the symmetry axis evolves near the compensation field, particularly in the presence of a slight misalignment~\cite{supp}.

In summary, we provide here a comprehensive semi-analyical picture for identifying the pseudo-OAM of azimuthal spin-waves and its practical implementation. The method relies on the fact in magnetic systems the SOI is controllable by the magnetic field and can be made very large. Our findings open a new avenue in the study of wave OAM by providing a robust detection mechanism not only for SWs but also potentially for photons and phonons through hybridization with magnons. The latter in particular offers a transduction of SAM to a mechanical torque, which could in principle be harnessed to rotate mesoscopic elastic bodies.

%\begin{figure}
%\includegraphics[width=0.45\textwidth]{FIG3_v4.png}
%\caption{\label{fig:3} Semi-analytical prediction of the evolution of the SOI between modes with $n_J\in [0,1,2]$ as a function of $\mu _0 M_s h_0$. The validity of the model ends below the softening of the $(0,0)$ mode at $H_\text{sat}$, which marks the onset of the cone state (see above texture).}
%\vspace{-10 pt}
%\end{figure}

\begin{acknowledgments}
 We sincerely thank Prof. Y. Otani from the University of Tokyo and Prof. C. Serpico from the University Federico II of Naples for their invaluable insights and thought-provoking discussions. This work was partially supported by the EU-project HORIZON-EIC-2021-PATHFINDER OPEN PALANTIRI-101046630; the French Grants ANR-21-CE24-0031 Harmony; the PEPR SPIN - MAGISTRAL ANR-24-EXSP-0004; the French Renatech network; and the REIMEI Research Program of Japan Atomic Energy Agency. K.Y. acknowledges support from JST PRESTO Grant No. JPMHPR20LB, Japan, JSPS KAKENHI (No. 21K13886), and JSPS Bilateral Program Number JPJSBP120245708. We also acknowledge partial support by the Japan Science and Technology Agency (JST) as part of Adopting Sustainable Partnerships for Innovative Research Ecosystem (ASPIRE), Grant Number JPMJAP2410.
\end{acknowledgments}

%\bibliography{AngularMomentum}% Produces the bibliography via BibTeX.


\end{document}

\title{Supplementary Material: "The Orbital Angular Momentum of Azimuthal Spin-Waves"}

\author{T. Valet \orcidlink{0000-0002-7775-9395}}
\email[Corresponding author :]{ tvalet@mphysx.com}
\affiliation{Université Grenoble Alpes, CEA, CNRS, Grenoble INP, Spintec, 38054 Grenoble, France}

\author{K. Yamamoto \orcidlink{0000-0001-9888-4796}}
\affiliation{Advanced Science Research Center, Japan Atomic Energy Agency, Tokai, Ibaraki 319-1195, Japan}

\author{B. Pigeau \orcidlink{0000-0003-2331-8436}}
\affiliation{Université Grenoble Alpes, CNRS, Grenoble INP, Institut Néel, Grenoble, France}

\author{G. de Loubens \orcidlink{0000-0001-8096-3058}}
\affiliation{SPEC, CEA, CNRS, Université Paris-Saclay, 91191 Gif-sur-Yvette, France}

\author{O. Klein \orcidlink{0000-0001-9429-5150}}
\email[Corresponding author :]{ oklein@cea.fr}
\affiliation{Université Grenoble Alpes, CEA, CNRS, Grenoble INP, Spintec, 38054 Grenoble, France}

\maketitle

\tableofcontents

% \clearpage
% \newpage

%%%%%%%%%%%%%%%%%%%%%%%%%%%%%%%%%%%%%%%%%%%%%%%%%%%%%%%%%%%%%%%%%%%%%%%%%%%%%%% 

\section{Sample}

A YIG thin film of nominal thickness 60~nm was grown on a 0.5~mm thick GGG(111) substrate by liquid phase epitaxy (LPE)~\cite{beaulieu2018}. The extended film was characterized by standard magnetometry, BLS, and broadband ferromagnetic resonance (FMR) techniques.

This YIG layer was subsequently patterned into a disk of nominal radius $R_{\rm disk} = 0.5$~$\mu$m using e-beam lithography and dry Ag etching. A 200~nm thick Ti/Au antenna of width 8~$\mu$m was then deposited on top of the disk. Injecting an rf current into it generates a linearly polarized in-plane rf magnetic field at the disk location. The output power of the rf synthesizer is set to $-30$~dBm, corresponding to an rf driving field of about 20~$\mu$T.

The exact thickness of the patterned disk was determined by BLS from the splitting between the Kittel mode and its first higher-order thickness standing spin-wave mode (exchange dominated). A value of $t_{\rm disk} = 55 \pm 2.5$~nm is extracted assuming an exchange length in YIG of $\lambda_{\rm ex} = 15$~nm~\cite{Klingler2014}.

A quite rough estimate of the saturation magnetization $M_s= 135 \pm 10$~kA/m is extracted by standard magnetometry, in agreement with previous studies on LPE films~\cite{beaulieu18}. We recall that the bulk value of $M_s$ of YIG at 300~K is 3.5 Bohr magneton per unit cell \cite{landolt70}, which translates into $M_s = 137 \pm 10$~kA/m . The accuracy of magnetometry of such thin film is limited primarily by the available sample volume, and not good enough for our purposes of quantitative comparison with the theory. A precise but not accurate evaluation of $M_s$ is inferred from the spectral position of the (0,1) mode measured by MRFM in the normal configuration. Its accuracy depends on too many other parameters: the disk radius, the disk thickness, the uniaxial anisotropy, the cubic anisotropy, the gyromagnetic ratio, the exchange constant, the tilt angle and the stray magnetic field of the MRFM probe. A better method would be to evaluate $M_s$ from the splitting between the (0,1) and (1,1) modes. Although the agreement between the model and the nominal value is excellent (see below) the uncertainty depends on how precise is our knowledge of the disk diameter, which ultimately is also not good enough for our purpose. 

A precise and somewhat more accurate method is the evaluation of the saturation magnetization extracted from FMR experiments performed on the continuous film, with the magnetic field applied in-plane along the $[11\overline{2}]$ axis. This eliminates the influence of the size of the disk, the tilt angle and the stray field of the MRFM, but not the anisotropies. Using the Kittel formula, we find the value of the effective magnetization $\mu_0 M_{\rm eff} = 0.1768\pm0.0002$~T (and the value of the gyromagnetic ratio given above). To extract an \emph{accurate} $\mu_0 M_s = \mu_0 (M_{\rm eff} + H_{K_1}/2 + H_{K_u})$~\cite{Makino1981}, one needs to remove the contributions of the first order cubic anisotropy, $\mu_0H_{K_1}  = -0.0085\pm0.0005$~T, and uniaxial anisotropy, $\mu_0 H_{K_u} = -0.002\pm0.002$~T, taken from literature. It is justified because the cubic anisotropy is independent of the YIG film thickness, while the uniaxial anisotropy is weak for LPE films in this thickness range~\cite{beaulieu18}. This leads to $\mu_0 M_s = 0.170\pm0.003$~T, in agreement with the expected value for LPE YIG thin films ($\mu_0 M_s = 0.17$~T, see~\cite{beaulieu18}).

Finally, the value of $\gamma$, the gyromagnetic ratio, is inferred with an accuracy better than 1\% from the frequency dependence of FMR in the saturated state, between 2 and 20 GHz. We obtain $\gamma = (1.77 \pm 0.01) \times 10^{11}~\mathrm{rad}\,\mathrm{s}^{-1}\,\mathrm{T}^{-1}$, in agreement with the expected value for YIG~\cite{beaulieu18}.  The Gilbert damping of the YIG film is $\alpha = 8 \times 10^{-5}$. 

%%%%%%%%%%%%%%%%%%%%%%%%%%%%%%%%%%%%%%%%%%%%%%%%%%%%%%%%%%%%%%%%%%%%%%%%%%%%%%% 

\section{Magnetic Resonance Force Microscopy}

The spin-wave mode spectroscopy of the YIG disk is performed by magnetic resonance force microscopy (MRFM)~\cite{naletov:2011}. It employs a very soft cantilever (spring constant $k=6$~mN/m), at the end of which a cobalt spherical probe of diameter 500~nm is attached~\cite{sangiao17}, to mechanically detect the magnetization dynamics in the sample placed underneath~\cite{klein08}. Using a piezo-electric displacement stage, this MRFM probe is set precisely above the center of the YIG disk to maintain the axial symmetry. The separation between the sample and the center of the probe is equal to 1.9~$\mu$m~\cite{Charbois2002}. When SWs are excited in the sample by the microwave field, the (static) longitudinal component of magnetization is reduced, hence the dipolar force on the MRFM probe. It results in a displacement of the cantilever beam, which is detected optically. The rf excitation applied to the sample via the antenna is pulse modulated at the mechanical resonance frequency $f_c \simeq 12.8$~kHz of the cantilever to improve the signal-to-noise ratio by its quality factor ($Q \simeq 2000$ under vacuum). A piezoelectric bimorph glued on the cantilever holder and a phase-locked loop are used to constantly drive the cantilever at resonance and fixed amplitude of 10~nm. The voltage applied on the bimorph corresponds to the MRFM signal. 

In our MRFM, one can align the disk normal with the axis of the electromagnet by introducing small wedges below the microscope, which allow to adjust the pitch and roll angle separately. The alignment is achieved with an accuracy better than $1^\circ$. The MRFM signal is then recorded while the out-of-plane applied magnetic field is slowly swept downwards from 0.2 to 0~T at constant rf frequency, which is varied by steps of 10~MHz from 1.9~GHz down to 10~MHz. Supplementary Figure~\ref{fig:comparison}(a) shows the measured spectra. In the following we will compare it with semi-analytical and numerical models.

%%%%%%%%%%%%%%%%%%%%%%%%%%%%%%%%%%%%%%%%%%%%%%%%%%%%%%%%%%%%%%%%%%%%%%%%%%%%%%% 
% \begin{figure}[t!]
% \includegraphics[width=0.9\textwidth]{211015b-1-pk.py.png}
% \caption{\label{fig:1} a) MRFM measurement of the Kittel resonance as a function of the applied magnetic field for increasing microwave excitation frequencies. b) Shift of the Kittel mode resonance frequency as a function of the applied field. The solid line shows a linear fit whose slope yields the gyromagnetic ratio. c) Variation of the full width at half maximum of the resonance peak as a function of the excitation frequency. The dashed line shows a linear fit whose slope indicates the Gilbert damping parameter.}
% % \vspace{-10 pt}
% \end{figure}

\section{Numerical simulation}

Our quantitative estimation of the SOI is based on the measurement of the frequency splitting between two eigenmodes. We show that this differential approach, relying on the frequency spacing between two spectral lines, is robust against external perturbations in particular a tilt angle. In the experiment a slight tilt $\theta_H$ of the magnetic field with respect to the normal seems necessary to excite the azimuthal modes using the spatially uniform rf field produced by the microwave antenna. In the following we aim to demonstrate that the impact of this tilt is anecdotal on the quantitative evaluation of the SOI because the whole spectrum shifts rigidly in the saturated state.

\begin{figure*}[t!]
\includegraphics[width=0.9\textwidth]{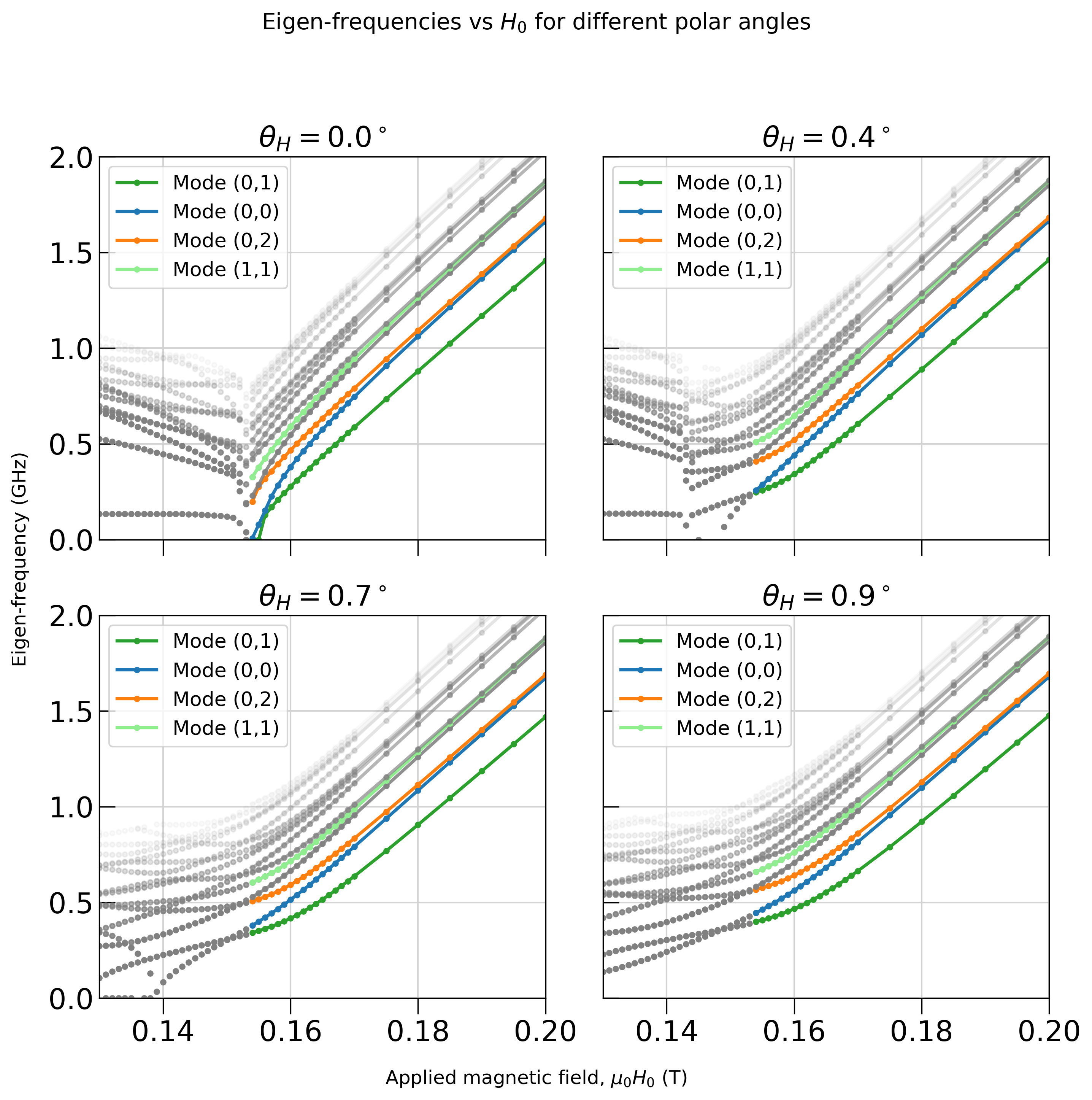}
\caption{\label{fig:3D} Numerical simulations using a home-developed  modal solver by finite element method showing the evolution of the lowest-energy modes as a function of the externally applied magnetic field $\mu_0 H_0$, oriented at polar angles of 0.0$^\circ$, 0.4$^\circ$, 0.7$^\circ$, and 0.9$^\circ$ with respect to the normal of the disk. The Kittel mode is labeled here $n_J=+1$ (green). The SOI splitting occurs between modes $n_J=0$ (orange) and $n_J=+2$ (blue).
}
% \vspace{-10 pt}
\end{figure*}

We have performed numerical calculations of the evolution with an externally applied magnetic field of the eigenfrequencies of the modes inside a disk of diameter $2 R_{\rm disk} = 1$~$\mu$m and thickness $t_{\rm YIG} = 55$~nm for four different values of the tilt angle, $\theta_H=$0, 0.4, 0.7, 0.9$^\circ$ between the applied magnetic field and the normal to the disk. These calculations were carried out using a custom modal solver, whose characteristics have been described elsewhere~\cite{Taurel2016}. Supplementary Figure~\ref{fig:3D} shows the corresponding deformation of the low energy spectrum. 

\begin{figure*}[t!]
\includegraphics[width=0.5\textwidth]{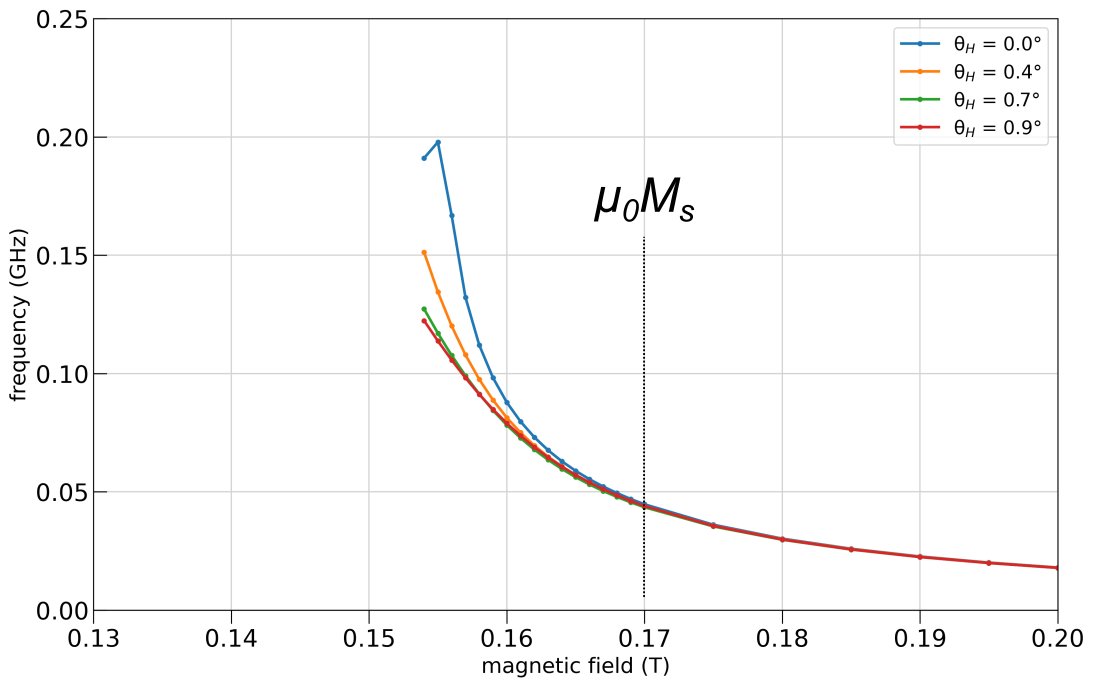}
\caption{\label{fig:angle} Extraction of the SOI splitting occurring between modes $n_J=0$ (orange) and $n_J=+2$ (blue) in FIG.\ref{fig:3D} for different values of $\theta_H$. 
}
% \vspace{-10 pt}
\end{figure*}

The SOI splitting occurs between modes (0,0) and (0,2). Supplementary Figure~\ref{fig:angle} shows the SOI extracted from the numerical simulations as a function of $H_0$ for the different values of the polar angle, $\theta_H$. The purpose of this plot is to show that SOI in the saturated state is indenpendent of $\theta_H$. This follows from the fact that the FMR spectrum of a saturated disk shifts rigidly with the tilt angle. Consequently, the value of the SOI, extracted from the separation between two spectral lines, is expected to remain independent of the tilt when $H_0 \geq M_s$. Concerning the deviations observed when $H_0 < M_s$, they are expected, since the magnetic susceptibility increases in the vicinity of the phase transition from the uniform state to a vortex.

\section{Quantitative comparison}

The purpose of this section is to corroborate the validity of our parameter values, which were determined through independent reasoning, against the main MRFM data. In Fig.~6 of the accompanying paper~\cite{valet:2024}, we have calculated the expected spectral position of the SW modes of the disk calculated at $\theta_H=0^\circ$ without taking into account the very small uniaxial anisotropy. The value of $\mu _0 M_s=0.170$~T used in the numerical model is the one of LPE YIG thin films~\cite{beaulieu18}. We provide in Supplementary Figure~\ref{fig:comparison} a new plot comparing Fig.~6 of the accompanying paper with the experimental spectra using the same scale. The value of the magnetic parameters and their uncertainty are summarized in Table~\ref{tab:param}. At $\mu_0 H_0=0.19$~T, we see that the experimentally measured 400~MHz splitting between the Kittel mode (0,1) and its first radial harmonics (1,1) matches the prediction indicating that both $R_{\rm disk}$ and $M_s$ are close to their nominal value. It also shows that a rigid shift could be used to match the experimental data. We now show on the same scale the numerical simulation at $\theta_H=0.7^\circ$. We see that the agreement between the model and the experimental data has much improved. Quantitatively, at $\mu_0 H_0=0.19$~T, the expected position of the (0,1) mode has been shifted from 1.1630~GHz at $\theta_H=0^\circ$ to 1.1866~GHz at $\theta_H=0.7^\circ$. The experimental value is at 1.279~GHz, thus 94~MHz higher in frequency, corresponding to a shift of about 0.0032~T. Estimating to $0.001$~T the stray magnetic field of the MRFM probe placed 1.9~$\mu$m above the disk, this leaves the remaining $0.002$~T to the uniaxial anisotropy, in agreement with previous studies~\cite{beaulieu18}. 

\begin{figure*}[t!]
\includegraphics[width=0.99\textwidth]{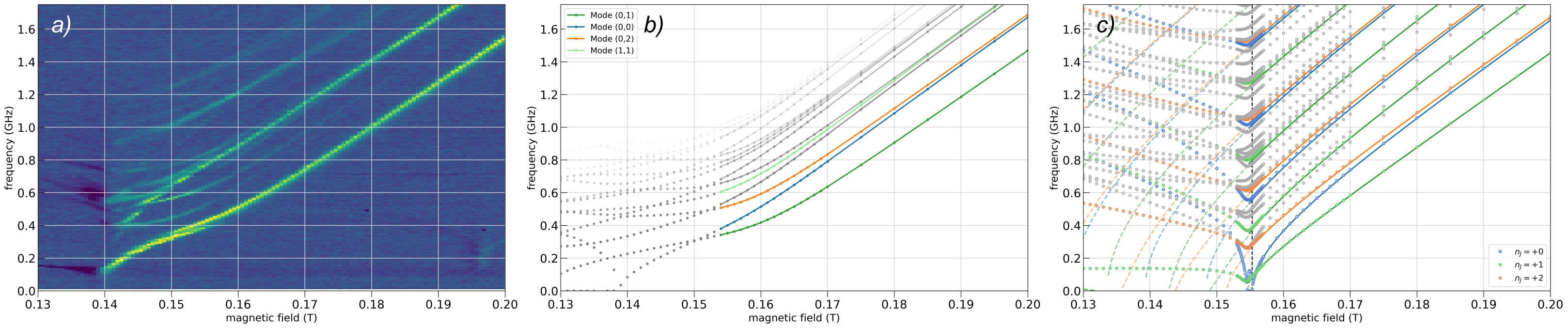}
\caption{\label{fig:comparison} Comparison on the same scale of the SW spectra (a) measured experimentally, (b) calculated numerically with $\theta_H=0.7^\circ$ and (c) calculated semi-analytically with  $\theta_H=0^\circ$ using the model described in Ref.~\cite{valet:2024}. The semi-analytical results are the solid lines. We have superimposed to (c) with dots the numerical values obtained by an axisymmetric solver using the same parameters. The axi-solver allows to add in the plot SW having $|n_J|>2$ as well to show the evolution for $H_0 < M_s$.
}
% \vspace{-10 pt}
\end{figure*} 

\begin{table}
  \caption{Magnetic parameters of the YIG disk: thickness, radius, exchange length, saturation magnetization, uniaxial anisotropy and gyromagnetic ratio.}
  \begin{ruledtabular}
    \begin{tabular}{c c c c c c}
      $t_{\rm disk}$  & $R_{\rm disk}$ & $\Lambda_{\rm ex}$ & $\mu_0 M_s$ &  $\mu_0 H_{Ku}$  &  
      $\gamma$  \\
     (nm)  &  (nm) & (nm) &  (T) & (T) & (rad.s$^{-1}$.T$^{-1}$) \\ \hline \\
      55 & 500 & 15 & $0.170$ &  0.002 & 
      $1.77 \cdot 10^{11}$  \\
      $\pm 2.5$ & $\pm 25$ & & $\pm 0.003$ & $\pm 0.002$ & $\pm 0.01$ \\
    \end{tabular}
  \end{ruledtabular}\label{tab:param}
\end{table}

\clearpage
\newpage
\section*{References}

%% %% Include extra code to remove duplicate TOC entry
%% \let\oldaddcontentsline\addcontentsline% Store \addcontentsline
%% \renewcommand{\addcontentsline}[3]{}% Make \addcontentsline a no-op
%% \bibliography{AngularMomentum}
%% \let\addcontentsline\oldaddcontentsline% Restore \addcontentsline